\documentclass[11pt,onecolumn,amssymb,nofootinbib]{revtex4}
\usepackage{bm}
\usepackage{amsmath}
\usepackage{amssymb}
\usepackage{graphicx}
\usepackage{amsmath}
\usepackage{bbm}

\begin{document}

\title{On Kalamidas'  proposal of faster than light quantum communication}

\author{GianCarlo Ghirardi}
\email{ghirardi@ictp.it}
\affiliation{Emeritus, University of Trieste\\
The Abdus Salam ICTP,Trieste, Strada Costiera 11, 34014 Trieste, Italy}.\\
\author{Raffaele Romano}
\email{rromano@iastate.edu}
\affiliation{Department of Mathematics, Iowa State University, Ames, IA (USA)}

\begin{abstract}
\noindent In a recent paper, Kalamidas has advanced a new proposal of faster than light communication which has not yet been proved invalid. In this paper, by strictly sticking to the standard quantum formalism, we prove that, as all previous proposals, it does not work. \\

\noindent KEY WORDS: faster-than-light signalling.
\end{abstract}
\maketitle

\section{Introduction and preliminaries on the experimental set up}
The idea that quantum entanglement and quantum interactions with a part of a composite system allow faster than light communication has been entertained for quite a long time. All existing proposals have been shown to be unviable. For a general overview we refer the reader to   papers by Herbert~\cite{he}, Selleri~\cite{se}, Eberhard~\cite{eb}, Ghirardi \& Weber~\cite{gw}, Ghirardi, Rimini \& Weber~\cite{grw}, Herbert~\cite{he2}, Ghirardi (who has derived the no-cloning theorem just to reject the challenging proposal [6] by Herbert - see the document attached to ref [7]), and, more recently, by Greenberger~\cite{gr} and Kalamidas~\cite{ka}. A detailed analysis of the problem and the explicit refutation of all proposals excluding the one of Kalamidas appear in  the recent work by Ghirardi~\cite{ghir}.

In view of the interest of the subject and of the fact that a lively debate on the topic is still going on  we consider our duty to make rigorously clear  that the  proposal [9] is basically flawed.

We will not go into   details concerning the precise suggestion and will simply present a very sketchy description of the experimental set-up. The main point can be grasped by the following picture, taken from the paper by Kalamidas, depicting a   source S of entangled photons in precise modes which impinge on appropriate beam splitters $BS_{0}, BS_{a}$ and $BS_{b}$, the first one with equal transmittivity and reflectivity, the other two with (real) parameters $t$ and $r$ characterizing such properties. Finally, in the region at right, one can (or not at his free will) inject coherent photon states $|\alpha\rangle$ characterized by the  indicated modes:

\begin{figure}[h]
\begin{center}
\includegraphics[scale=0.6]{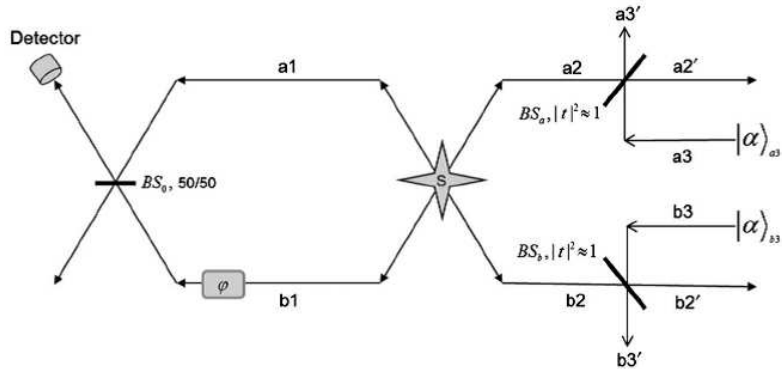}
\caption{The experimental set up devised by Kalamidas.} \label{f1}
\end{center}
\end{figure}

\section{The initial state}

Kalamidas' mechanism for superluminal signaling rests on the possibility of injecting or to avoid to do so the coherent states at the extreme right. Correspondingly, one has, as his initial state either:

\begin{equation}
|\psi_{in}\rangle=\frac{1}{\sqrt{2}}(a_{1}^{\dag}a_{2}^{\dag}+e^{i\phi}b_{1}^{\dag}b_{2}^{\dag})D_{a3}(\alpha)D_{b2}(\alpha)|0\rangle,
\end{equation}
\noindent where $D_{a3}(\alpha)$ and $D_{b3}(\alpha)$ are coherent states of modes $a3$ and $b3$, or, alternatively, the state:

\begin{equation}
|\tilde{\psi}_{in}\rangle=\frac{1}{\sqrt{2}}(a_{1}^{\dag}a_{2}^{\dag}+e^{i\phi}b_{1}^{\dag}b_{2}^{\dag})|0\rangle.
\end{equation}

\section{Second step: the evolution and the beam splitters}

Once one has fixed the initial state the process starts and the state evolves in time. The evolution implies the passage of photons through the indicated beam splitters. It has to be mentioned that the recent debate on Ref.[9] has seen disagreeing positions concerning the  functioning of these devices. 

We will not enter into  technical details, we   simply describe the effect of crossing a beam splitter in terms of appropriate unitary operations which account for its functioning. The result is  the one considered by Kalamidas. Let me stress, due to its importance, that this move to simply consider the unitary nature of the transformations overcomes any specific debate. Actually, the quite general and legitimate assumption that {\bf any unitary transformation of the Hilbert space can actually be implemented} makes useless  entering into the details of the functioning of the beam splitters, a move that we consider important since, apparently, different people make different claims concerning such a functioning. I simply consider the evolution of the initial statevector induced by the unitary transformation $U=U_{0}U_{a}U_{b}$, with:

\begin{eqnarray}
U_{0}a_{1}^{\dag}U_{0}^{\dag}&= &\frac{1}{\sqrt{2}}(a_{1}^{\dag}+b_{1}^{\dag}) \nonumber \\
U_{0}b_{1}^{\dag}U_{0}^{\dag}&=& \frac{1}{\sqrt{2}}(-a_{1}^{\dag}+b_{1}^{\dag}) \nonumber \\
U_{a}a_{2}^{\dag}U_{a}^{\dag}&=&(ta_{2}^{\dag}+ra_{3}^{\dag}) \nonumber \\
U_{a}a_{3}^{\dag}U_{a}^{\dag}&=&(-rb_{2}^{\dag}+tb_{3}^{\dag})\nonumber \\
U_{b}b_{2}^{\dag}U_{b}^{\dag}&=&(tb_{2}^{\dag}+rb_{3}^{\dag}) \nonumber \\
U_{b}b_{3}^{\dag}U_{b}^{\dag}&=&(-rb_{2}^{\dag}+tb_{3}^{\dag}).
\end{eqnarray}

Using such expressions one easily evaluates the evolved of each of the two initial states  going through all the beam splitters with their particular characteristics.The computation is quite easy and the final state, when the coherent states are present at  right, turns to have the following form:

\begin{eqnarray}
|\psi_{fin}\rangle &\equiv& U_{0}U_{a}U_{b}|\psi_{in}\rangle\nonumber \\
&=&\frac{1}{2}[(a_{1}^{\dag}+b_{1}^{\dag})(ta_{2}^{\dag}+ra_{3}^{\dag})+e^{i\phi}(-a_{1}^{\dag}+b_{1}^{\dag})(tb_{2}^{\dag}+rb_{3}^{\dag})]\nonumber \\
&\times& D_{a3}(t\alpha)D_{a2}(-r\alpha)D_{b3}(t\alpha)D_{b2}(-r\alpha)|0>.
\end{eqnarray}

\noindent Alternatively, when the second initial state is considered,  the evolution leads to:

 \begin{eqnarray}
|\tilde{\psi}_{fin}\rangle&\equiv & U_{0}U_{a}U_{b}|\tilde{\psi}_{in}\rangle \nonumber \\
&=& \frac{1}{2}[(a_{1}^{\dag}+b_{1}^{\dag})(ta_{2}^{\dag}+ra_{3}^{\dag})+e^{i\phi}(-a_{1}^{\dag}+b_{1}^{\dag})(tb_{2}^{\dag}+rb_{3}^{\dag})]|0\rangle.
\end{eqnarray}

\section{Possible actions at right}

I must confess that the original paper by Kalamidas as well as many of the comments which followed are not sufficiently clear concerning what one does at right on the photons  appearing there. One finds statements of the type ``when there is one photon in mode ${\bf a2'}$ and  one photon in mode ${\bf b2'}$" then ``there is a coherent superposition of single photon occupation possibilities between modes ${\bf a1}$ and ${\bf b2}$". Here I cannot avoid stressing that such statements, as they stand, are meaningless because they take into account one of the possible outcomes and not the complete unfoldng of the measurement process. If one is advancing a precise proposal for an experiment, he must clearly specify which actions are actually performed. And here comes the crucial point: the  alleged important consequences of an action performed at right on the outcomes at left must be deduced from the analysis of the outcomes of possible observations in the region at left (we want to have a signal there). It seems to me that the proponent of the new mechanism for superluminal communication has not taken into account a fundamental fact which has been repeatedly stressed precisely in the literature on the subject. What we have to investigate are the implications of precise actions at right for the physics of the systems in the region at left. In turn, all what is physically relevant at left,  as well known, is exhaustively accounted by the reduced statistical operator $\rho_{red}^{(L)}$ referring to the systems which are there, i.e. the one obtained from the full statistical operator $\rho(L,R)\equiv |\psi_{fin}\rangle\langle\psi_{fin}|$ by taking the partial trace on the right degrees of freedom: $\rho_{red}^{(L)}=Tr^{(R)} [ |\psi_{fin}\rangle\langle\psi_{fin}|]$, with obvious meaning of the symbols. Now, the  operator $\rho_{red}^{(L)}$ is unaffected by all conceivable legitimate actions made at right. The game is the usual one. One can consider:

\begin{itemize}
\item Unitary evolutions involving the systems at right : $\rho(L,R)\rightarrow U^{(R)}\rho(L,R)U^{\dag(R)}$
\item Projective measurement of an observable with spectral family $P_{k}^{(R)}: \rho(L,R)\rightarrow\sum_{k}P_{k}^{(R)}\rho(L,R)P_{k}^{(R)}$
\item Nonideal measurements associated to a family $A_{k}^{(R)}: \rho(L,R)\rightarrow\sum_{k}A_{k}^{(R)}\rho(L,R)A_{k}^{\dag(R)}$.
\end{itemize}

In all these cases (which exhaust all legitimate quantum possibilities), due to the cyclic property of the trace, to the unitarity of $U^{(R)}$ and to the fact that the projection operators $P_{k}^{(R)}$ as well as the quantities $A_{k}^{\dag(R)}A_{k}^{(R)}$  sum to the identity operator (obviously the one referring to the Hilbert space of the systems at right), the reduced statistical operator $\rho_{red}^{(L)}$ does not change in any way whatsoever as a consequence of the action at right. 

In brief, for investigating the physics at left one can ignore completely possible evolutions or measurements of any kind done at right. Obviously the same does not hold if one performs a selective measurement at right. But in this case the changes at left induced by the measurement depend on the  outcome which one gets, so that, to take advantage of the change, the receiver at left must be informed concerning the outcome at right, and this requires a luminal communication. In accordance with these remarks, sentences like those I have mentioned above and appearing in Ref.[8], must be made much more precise. If at right one performs a measurement identifying the occupation numbers of the various states, one has to describe it appropriately taking into account all possible outcomes. Concentrating the attention on a specific outcome one is actually considering a selective measurement, an inappropriate procedure, as just discussed.

Concluding this part:  to compare the situation at left in the case in which at right coherent states are injected or not, one can plainly work with the evolved states (4) and (5). The fundamental question concerning the possibility of superluminal communication becomes  then: does it exist an observable for the particles at left (i.e. involving modes ${\bf a1}$ and ${\bf b1}$) which has a different mean value or spread or probability for individual outcomes when the state is the one of Eq.(4) or the  one of Eq.(5)?

\section{Proof that no effect is induced at left}

In accordance with the previous analysis, to answer the just raised question  we consider  the most general self-adjoint operator of the Hilbert space of the modes at left which we will simply denote as $h(a_{1},a_{1}^{\dag},b_{1},b_{1}^{\dag})$, and we will evaluate its mean value in the two states (4) and (5). 

In the case of state (5), $|\psi_{fin}\rangle$ we have:

\begin{eqnarray}
& &\langle \psi_{in} |U^{\dag}h(a_{1},a_{1}^{\dag},b_{1},b_{1}^{\dag})U|\psi_{in}\rangle \equiv \nonumber \\
& &\frac{1}{4}\langle 0|D_{b3}^{\dag}(t\alpha) D_{b2}^{\dag}(-r\alpha)D_{a3}^{\dag}(t\alpha)D_{a2}^{\dag}(-r\alpha)[(a_{1}+b_{1})(t a_{2}+r a_{3})+e^{i\phi}(-a_{1}+b_{1})(t b_{2}+r b_{3})] h(a_{1},a_{1}^{\dag},b_{1},b_{1}^{\dag})\nonumber \\
& & [(a_{1}^{\dag}+b_{1}^{\dag})(t a_{2}^{\dag}+r a_{3}^{\dag})+e^{i\phi}(-a_{1}^{\dag}+b_{1}^{\dag})(t b_{2}^{\dag}+r b_{3}^{\dag})]D_{a2}(-r\alpha)D_{a3}(t\alpha)D_{b2}(-r\alpha)D_{b3}(t\alpha)|0\rangle.
\end{eqnarray}

One has now to take into account that the vacuum is the product of the vacua for all modes, $|0\rangle=|0\rangle_{1}|0\rangle_{2}|0\rangle_{3}$. The previous equation becomes:

\begin{eqnarray}
& &\langle \psi_{in} |U^{\dag}h(a_{1},a_{1}^{\dag},b_{1},b_{1}^{\dag})U|\psi_{in}\rangle \equiv \nonumber \\
& &\frac{1}{4}[ _{2}\langle0|_{3}\langle 0|D_{b3}^{\dag}(t\alpha) D_{b2}^{\dag}(-r\alpha)D_{a3}^{\dag}(t\alpha)D_{a2}^{\dag}(-r\alpha)(t a_{2}+r a_{3})(t a_{2}^{\dag}+r a_{3}^{\dag}) \nonumber \\
& &D_{a2}(-r\alpha)D_{a3}(t\alpha)D_{b2}(-r\alpha)D_{b3}(t\alpha)|0\rangle_{2}|0\rangle_{3}]\cdot _{1} \langle0|(a_{1}+b_{1})h(a_{1},a_{1}^{\dag},b_{1},b_{1}^{\dag})(a_{1}^{\dag}+b_{1}^{\dag})|0\rangle_{1}+\nonumber \\
& &\frac{1}{4}e^{i\phi}[ _{2}\langle0|_{3}\langle 0|D_{b3}^{\dag}(t\alpha) D_{b2}^{\dag}(-r\alpha)D_{a3}^{\dag}(t\alpha)D_{a2}^{\dag}(-r\alpha)(t a_{2}+r a_{3})(t b_{2}^{\dag}+r b_{3}^{\dag}) \nonumber \\
& &D_{a2}(-r\alpha)D_{a3}(t\alpha)D_{b2}(-r\alpha)D_{b3}(t\alpha)|0\rangle_{2}|0\rangle_{3}]\cdot _{1} \langle0|(a_{1}+b_{1})h(a_{1},a_{1}^{\dag},b_{1},b_{1}^{\dag})(-a_{1}^{\dag}+b_{1}^{\dag})|0\rangle_{1}+\nonumber \\
& &\frac{1}{4}e^{-i\phi}[ _{2}\langle0|_{3}\langle 0|D_{b3}^{\dag}(t\alpha) D_{b2}^{\dag}(-r\alpha)D_{a3}^{\dag}(t\alpha)D_{a2}^{\dag}(-r\alpha)(t b_{2}+r b_{3})(t a_{2}^{\dag}+r a_{3}^{\dag}) \nonumber \\
& &D_{a2}(-r\alpha)D_{a3}(t\alpha)D_{b2}(-r\alpha)D_{b3}(t\alpha)|0\rangle_{2}|0\rangle_{3}]\cdot _{1} \langle0|(-a_{1}+b_{1})h(a_{1},a_{1}^{\dag},b_{1},b_{1}^{\dag})(-a_{1}^{\dag}+b_{1}^{\dag})|0\rangle_{1}+\nonumber \\
& &\frac{1}{4}[ _{2}\langle0|_{3}\langle 0|D_{b3}^{\dag}(t\alpha) D_{b2}^{\dag}(-r\alpha)D_{a3}^{\dag}(t\alpha)D_{a2}^{\dag}(-r\alpha)(t b_{2}+r b_{3})(t b_{2}^{\dag}+r b_{3}^{\dag}) \nonumber \\
& &D_{a2}(-r\alpha)D_{a3}(t\alpha)D_{b2}(-r\alpha)D_{b3}(t\alpha)|0\rangle_{2}|0\rangle_{3}]\cdot _{1} \langle0|(-a_{1}+b_{1})h(a_{1},a_{1}^{\dag},b_{1},b_{1}^{\dag})(-a_{1}^{\dag}+b_{1}^{\dag})|0\rangle_{1}.
\end{eqnarray}

Let us take now into consideration the expression in square brackets of the first term (the one which contains the coherent states and the vacua for modes 2 and 3). If one keeps in mind that the coherent states are eigenstates of the annihilation operators one can apply the four terms arising from the expression  $(t a_{2}+r a_{3})(t a_{2}^{\dag}+r a_{3}^{\dag})$ to the coherent states. Obviously, before doing this one has to commute the operators $a_{2}$ and $a_{2}^{\dag}$ in the expression $ta_{2}a_{2}^{\dag}$ and the similar one for mode 3. In so doing the expression $(t a_{2}+r a_{3})(t a_{2}^{\dag}+r a_{3}^{\dag})$ reduces to 1. Just for the same reason and with the same trick one shows that one can replace with 1  the expression $(t b_{2}+r b_{3})(t b_{2}^{\dag}+r b_{3}^{\dag})$ in the last term. The same calculation shows also that the corresponding expressions in the secon and third terms reduce to 0. The final step consist therefore in evaluating, for the first and fourth terms the expressions:

\begin{equation}
[ _{2}\langle0|_{3}\langle 0|D_{b3}^{\dag}(t\alpha) D_{b2}^{\dag}(-r\alpha)D_{a3}^{\dag}(t\alpha)D_{a2}^{\dag}(-r\alpha)
D_{a2}(-r\alpha)D_{a3}(t\alpha)D_{b2}(-r\alpha)D_{b3}(t\alpha)|0\rangle_{2}|0\rangle_{3}].
\end{equation}

Taking into account that $D_{a}^{\dag}(\alpha)D_{a}(\alpha)=I$ one gets the final expression for the expectation value of the arbitrary hermitian operator $h(a_{1},a_{1}^{\dag},b_{1},b_{1}^{\dag})$ when one starts with the initial state containing the coherent states:

\begin{eqnarray}
& &\langle \psi_{in} |U^{\dag}h(a_{1},a_{1}^{\dag},b_{1},b_{1}^{\dag})U|\psi_{in}\rangle =\nonumber \\
& &\frac{1}{4}[_{1} \langle0|(a_{1}+b_{1})h(a_{1},a_{1}^{\dag},b_{1},b_{1}^{\dag})(a_{1}^{\dag}+b_{1}^{\dag})|0\rangle_{1}+\nonumber \\
& & _{1} \langle0|(-a_{1}+b_{1})h(a_{1},a_{1}^{\dag},b_{1},b_{1}^{\dag})(-a_{1}^{\dag}+b_{1}^{\dag})|0\rangle_{1}].
\end{eqnarray}

It is now an easy game to repeat the calculation for the much  simpler case in which the initial state is $|\tilde{\psi}\rangle$. One simply has precisely the expression (7) with all the coherent states missing. Taking into account that the operators of modes 2 and 3 act now on the vacuum state one immediately realizes that one gets once more the result  (9).

\section{Conclusion}
We have proved, with complete rigour that the expectation value of any conceivable self adjoint operator of the space of the modes 1 at left remains the same when one injects or does not inject the coherent states at right. Note that the result is completely independent from the choice of the phase $\phi$ characterizing the two terms of the entangled initial state and from the parameters $t$ and $r$ of the beam splitters and it does not involve any approximate procedure.

Accordingly, we have shown once more  that devices of the type of the one suggested by Kalamidas do not consent superluminal communication.

 A last remark. During the alive debate which took place recently in connexion with Kalamidas proposal, other authors have reached the same conclusion. However the reasons for claiming this were not always crystal clear and a lot of discussion had to do with the approximations introduced by Kalamidas. For these reasons we have decided to be extremely general and we have been pedantic in discussing even well known facts and properties of an ensemble of photons. Our aim has been to  refute in a completely clean and logically consistent way the idea that the device consents faster than light signaling.

\end{document}